# Exploration and Comparison of Deep Learning Architectures to Predict Brain Response to Realistic Pictures

Riccardo Chimisso*[1], Sathya Buršić[1], Paolo Marocco, Giuseppe Vizzari[1], Dimitri Ognibene[1,2]*

[1]Università degli Studi di Milano Bicocca, Milan, Italy
[2]University of Essex, Colchester, UK

*Corresponding authors: r.chimisso@campus.unimib.it, dimitri.ognibene@unimib.it

## Table of contents

# Abstract

We present an exploration of machine learning architectures for predicting brain responses to realistic images on occasion of the Algonauts Challenge 2023.

Our research involved extensive experimentation with various pretrained models. Initially, we employed simpler models to predict brain activity but gradually introduced more complex architectures utilizing available data and embeddings generated by large-scale pre-trained models. We encountered typical difficulties related to machine learning problems, e.g. regularization and overfitting, as well as issues specific to the challenge, such as difficulty in combining multiple input encodings, as well as the high dimensionality, unclear structure, and noisy nature of the output.

To overcome these issues we tested single edge 3D position-based, multi-region of interest (ROI) and hemisphere predictor models, but we found that employing multiple simple models, each dedicated to a ROI in each hemisphere of the brain of each subject, yielded the best results - a single fully connected linear layer with image embeddings generated by CLIP as input. While we surpassed the challenge baseline, our results fell short of establishing a robust association with the data.

# 1. Introduction

## 1.1. Challenge

The Algonauts Project[2], first launched in 2019, is on a mission to bring biological and machine intelligence researchers together on a common platform to exchange ideas and pioneer the intelligence frontier. The Algonauts 2023 Challenge[4] focuses on predicting responses in the human brain as participants perceive complex natural visual scenes. Through collaboration with the Natural Scenes Dataset (NSD)[1] team, the Challenge runs on the largest suitable brain dataset available, opening new venues for data-hungry modeling.

## 1.2. Dataset

The NSD dataset[1] consists of whole-brain, high-resolution fMRI[13] measurements in ultra-high-field (7T) strength of 8 healthy adult subjects while viewing thousands of natural scenes over the course of 30-40 scan sessions.

Each subject viewed 10,000 images, with 1000 shared across all subjects. In total 73,000 images (8 subjects times 9000 unique images plus 1000 shared ones). Each image was presented 3 times, meaning 30,000 image trials per subject.

Subjects were instructed to focus on a fixation cross at the center of the screen and perform a continuous recognition task in which they reported whether the current image had already been presented.

The fMRI data of the last three sessions of every subject is withheld and constitutes the basis for the test split. The fMRI data is divided into left/right hemispheres, each with 19k/20k vertices. Subjects 6 and 8 have less due to missing data.

Voxel activity is z-scored for each session, and then the fMRI responses are averaged across repeats of the same stimulus.

The NSD images are cropped versions of the COCO dataset[13] images.

## 1.3. Tested Input Pretrained Embeddings

To allow for greater efficiency and leverage transfer learning, we adopted state-of-the-art networks to create embeddings providing input representations that would facilitate generalization. In particular we explored the use of:

- **AlexNet**
  AlexNet[9] is a CNN architecture that has played a significant role in the advancement of machine learning and computer vision. PyTorch provides a pretrained version of AlexNet, used in the challenge development kit. Features from its layers (`features.2`) are extracted.
- **Clip**
  CLIP (Contrastive Language-Image Pretraining)[10] is a powerful model that can understand images, text, and their relationships. Two transformers, a vision and a text one, are trained jointly with the aim of maximizing similarity between transformations of related images and texts while minimizing it for unrelated pairs. CLIP can create embeddings small yet rich in semantics.
- **SAM**
  SAM (Segment Anything Meta)[8] aims to create valid segmentation masks given any prompt for any image. Similarly to NLP models, it's expected to output a coherent response to ambiguity: generating a valid mask means that even when a prompt could refer to multiple objects, the output should be a reasonable mask for at least one of them.
  SAM uses a fine-tuned pre-trained MAE[6] (Masked AutoEncoder) ViT[3] (Vision Transformer) which runs only once for each image. The resulting embeddings are large (64, 64, 256) but provide a compact representation of the visual content of the image.

## 1.4. Hardware

The available hardware consisted of a remote machine with:
- Intel Core i9-13900K 24c (32t) up to 5.80GHz with 32MB L2 cache and 125W TDP
- 2x NVidia GPUs RTX 3090 Ti 10752 CUDA cores up to 1.86GHz, with 24GB VRAM, 384-bit bus, and 450W TDP
- 128GB RAM

# 2. Approaches

The challenge baseline consisted of a linear regression model fit on features extracted from a layer of AlexNet, one for each hemisphere of each subject.

After searching for other models that could provide more semantically rich embeddings, we approached the challenge with simple models such as single linear layers and DNNs trained on some of the embeddings we had at our disposal provided by the selected large-scale pre-trained models.

We faced many challenges in trying to overcome early overfitting of our models, where only the simplest models managed to achieve good loss before overfitting. We applied different regularization techniques, such as dropout, batch normalization, weight decay, learning rate decay, etc., but to little or no avail. Furthermore, the data for the challenge is inherently noisy and not as large as other CV datasets. Moreover, AlexNet and SAM provided large embeddings that were difficult to handle efficiently with the available hardware and small dataset.

We attempted to devise more complex and powerful models, designed also to reduce the dimensionality of the aforementioned embeddings, with some even taking as input carefully constructed positional encodings of the 3D coordinates of the node whose activity they aim to predict.

To find the right balance between regularization and specialization, we also differentiated the models by focus: predicting a whole hemisphere at a time, single ROI or single vertex, as well ROIs pairs or groups from the two hemispheres; trained single-subject or cross-subject.

At the moment we found that employing multiple simple models, each dedicated to a ROI in each hemisphere of the brain of each subject, yielded the best results. This may also be due to higher training times and number of hyperparameters to explore for more complex models.

By specializing our models, we make them predict more specific data and thus have better chances to find a pattern with the available data. For instance, we could consider a model that predicts a single ROI more specialized than one that predicts a whole hemisphere. However, this may also lead to an increase in overfitting, and that's where more or less aggressive regularization strategies come in hand.

We thus divide our attempts into three main categories detailed below.

## 2.1. Linear

Our first experiment consisted of taking the same baseline model, but instead of feeding the linear regression with features from AlexNet, we fed embeddings extracted from CLIP. CLIP has the advantage of creating embeddings very rich in semantics despite being only a single tensor of shape 512.

This was unexpectedly better than we thought it would be, with 5 ROIs above the 0.5 correlation threshold and an average correlation of about 0.4, more than a 0.1 increase over the linear regression based on AlexNet. It's important to note, however, that despite the average being higher, some ROIs faced a decrease in their correlation score.

This suggests that, while CLIP's embeddings are better overall, for some specific areas, AlexNet's embeddings appear to be more suitable, maybe due to the latter being more sensitive to the spatial and visual characteristics of the image, whereas CLIP embeddings mainly represent semantics.

Since all our further attempts to improve the performance of our best model weren't successful, we tried a different approach by going back to the starting point and trying to focus more on the prediction of the best model to improve its performance.

In practice, what we did was apply the same model architecture as our best model, a single linear layer with input CLIP embeddings, but this time it would predict all vertex activations of a single ROI at a time: input CLIP, output N activations, with N being the size of the ROI.

Surprisingly, this model did improve performance compared to our previous best one; not by a wide margin, but it still improved the correlation scores across the board by more than 0.01. We relate this improvement to the behavior of the optimizer and to the possibility of adopting separate early stopping adopted for each ROI.

## 2.2. Deep Models

To avoid dealing with too many network parameters the dimensions of the embeddings have to also be modest. This unfortunately excluded SAM and AlexNet, and left only CLIP as image feature generator. We proceeded to create a new fully connected Deep Neural Network architecture for it, namely 4 linear blocks and one linear layer, where a linear block consisted of a sequence of 3 layers: Linear, BatchNorm1d[7] and LeakyReLU.

The numerical data that we produced proved that this architecture, while also taking more time to train and evaluate, given its higher complexity, was also producing on average slightly poorer results and overfitting. With the same architecture we trained the model on data of all subjects, and this time too we found that generalizing across multiple subjects isn't trivial.

To try and further improve the results we obtained with the previous experiments we introduced dropout, since the model was starting to get deep, and we also attempted to introduce some residual connection[5] blocks in the model architecture.

We kept the same 5% dropout in the other non-residual blocks. To increase the capacity of the model, we increased the size of the layer at each non-residual block by a factor of 1.25.

Unfortunately, our best configuration with this architecture still performed worse than the original one, although it did overfit less. The correlation scores were similar, along with the minimum values of both train and validation losses. However, this model was more stable, most likely due to the residual blocks allowing a better propagation of the gradient from the last layers, the ones closer to the output, to the first layers, the ones closer to the input.

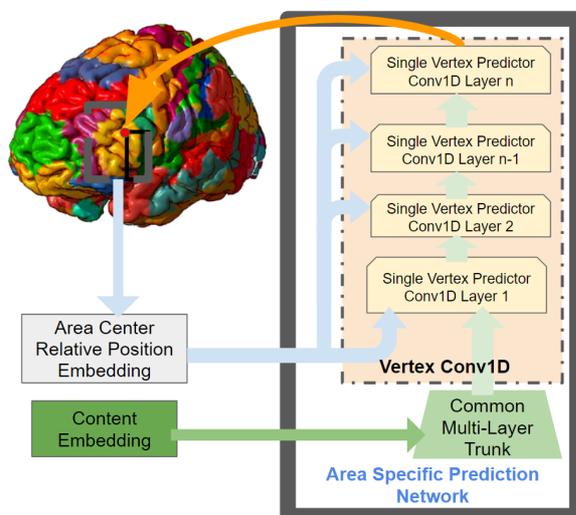

## 2.3. Single Edge Area Specific Positional Encoding

The positional encoding architecture had a double aim: first, to reduce the dimensionality of the output layer; second, to provide spatial information on the output edge. This information was not available to fully connected layers, but it could be important as often close-by vertices may show similar responses. Deconvolution layers would instead induce translation properties that don't fit the task at hand.

As shown in the above figure, the model was divided in two components: first a compressive Multi-Layer Trunk (which was Conv2d with SAM embeddings as input) that processed the input embedding producing an area specific representation. This component could process substantially bigger embeddings even with limited computational resources because it was shared between all the vertices of the ROI while its output was independent of their number. The second part would predict in parallel the activation of each vertex by performing a convolution on a vector where each element corresponded to a vertex. Each element of the vector was the concatenation of the area specific representation and the relative 3D position of the corresponding vertex within the current ROI using for each dimension a positional encoding similar to that adopted in transformers[11].

To test the capability of our settings, we initially checked if the model was able to reconstruct the actual 3D position. While we tested a few configurations, the model was always overfitting or underfitting. Further experimentation with the hyperparameters would be necessary to better understand this architecture.

# 4. Results

| $\beta_1$ | $\beta_2$ | $\varepsilon$ | Weight decay[14] |
|---|---|---|---|
| 0.9 | 0.999 | 1e-8 | 1e-2 |

All models used the AdamW[17] optimizer with default parameters reported on the left. During training and validation we used an MSE loss as it provided more stable results.

All model metrics refer to subject 01 and to the OFA ROI, unless the model was predicting all ROIs simultaneously; in that case, the minimum loss values are averaged across all ROIs, however the correlation values will still be about OFA. Training loss, validation loss and correlation values are separately presented for the left (above) and right (below) hemispheres. We report the best results for each of the described configurations.

| Experiment | Focus | Config | Architecture | T-loss | V-loss | Corr |
|---|---|---|---|---|---|---|
| CLIP-Linear | All ROIs | LR: 1e-03<br>Decay Step: 10<br>Decay: 0.5 | Single linear layer | 0.3821 | 0.4214 | 0.2726 |
| | | | | 0.3784 | 0.4134 | 0.3257 |
| CLIP-Linear | Single | LR: 1e-04<br>Decay Step: 10<br>Decay: 0.75 | Single linear layer | 0.4316 | 0.4711 | **0.2747** |
| | | | | 0.4104 | 0.4517 | **0.3274** |
| CLIP-DM | Single | LR: 5e-06<br>Decay Step: 10<br>Decay: 0.75 | 6 Conv1d layers, BatchNorm1d and LeakyReLU, **positional "attention"** at each Conv1d layer | 0.3268 | 0.4946 | 0.2157 |
| | | | | 0.3153 | 0.4687 | 0.2835 |
| CLIP-Pos (Dropout) | Single | LR: 7.5e-06<br>Decay Step: 5<br>Decay: 0.75 | Trunk=identity<br>Vertex Conv1D: 6 Conv1d layers, BatchNorm1d and LeakyReLU, **positional input** at each Conv1d layer, **20% dropout** every 2 Conv1d layers | 0.4529 | 0.4809 | 0.2528 |
| | | | | 0.4311 | 0.4596 | 0.3180 |
| CLIP-Pos (Resblocks) | Single | LR: 1.5e-05<br>Decay Step: 5<br>Decay: 0.75 | Trunk=identity<br>Vertex Conv1D: 6 Conv1d layers, BatchNorm1d and LeakyReLU, **positional input** at each Conv1d layer, **20% dropout** and a **resblock** every 2 Conv1d layers | 0.3846 | 0.4885 | 0.2545 |
| | | | | 0.3747 | 0.4667 | 0.3125 |

## 5. Conclusion

In our experiments the simpler architectures proved to be more robust and offer higher performance. On the other hand, more complex models, while offering greater flexibility and showing better results during training, took longer to fine-tune and didn't perform well on the validation dataset. We aim to test several other methods to improve their performance. The positional model could have exploited transformer-based layers as well as test and pre-training using artificial datasets using 3D grids with different outputs corresponding to different 3d positions and embeddings. This approach could help us gain a deeper understanding of how the system behaves and enable us to fine-tune its settings more effectively through targeted optimization of hyperparameters. In our future research, we'll explore the integration of various regularization techniques, like L1 regularization. This would allow us to combine different base models and embeddings that are better suited for different aspects of the image, such as its spatial and semantic features. Additionally, we didn't take into account certain specific details, such as the order in which the images were presented and the potential impact of attention due to image saliency[15,16]]. Given that subjects viewed some images multiple times, this information might have played a significant role in our results.